\documentclass[prd,preprint,tightenlines,showpacs,preprintnumbers,nofootinbib,eqsecnum,superscriptaddress]{revtex4-1}

 \usepackage[dvips,final]{graphicx}
  \usepackage{amssymb}
   \usepackage{amsmath}
    \usepackage{amsfonts}
     \usepackage{epsfig}
      \usepackage{bm}

\usepackage[section]{placeins}

\usepackage{sidecap}
\usepackage{multirow}
\usepackage{booktabs}
\usepackage{array}
\usepackage{tabularx}
\usepackage{xcolor}
\usepackage{pstricks}
\usepackage{diagbox}

\begin{document}

\begin{flushright}
MS-TP-23-24
\end{flushright}
\newcommand{\WS}[1]{{\red WS: #1}}
\newcommand{\VPG}[1]{{\blue VPG: #1}}

\newcommand{\ds}{\displaystyle}
\newcommand{\mc}{\multicolumn} 
\newcommand{\bce}{\begin{center}}
\newcommand{\ece}{\end{center}}
\newcommand{\beq}{\begin{equation}}
\newcommand{\eeq}{\end{equation}}
\newcommand{\bea}{\begin{eqnarray}}

\newcommand{\eea}{\end{eqnarray}}
\newcommand{\cont}{\nonumber\eea\bea}
\newcommand{\cl}[1]{\begin{center} {#1} \end{center}}
\newcommand{\ea}{\end{array}}

\newcommand{\ab}{{\alpha\beta}}
\newcommand{\cd}{{\gamma\delta}}
\newcommand{\dc}{{\delta\gamma}}
\newcommand{\ac}{{\alpha\gamma}}
\newcommand{\bd}{{\beta\delta}}
\newcommand{\abc}{{\alpha\beta\gamma}}
\newcommand{\eps}{{\epsilon}}
\newcommand{\lam}{{\lambda}}
\newcommand{\mn}{{\mu\nu}}
\newcommand{\mpnp}{{\mu'\nu'}}
\newcommand{\Amuu}{{A_{\mu}}}
\newcommand{\Amuo}{{A^{\mu}}}
\newcommand{\Vmuu}{{V_{\mu}}}
\newcommand{\Vmuo}{{V^{\mu}}}
\newcommand{\Anuu}{{A_{\nu}}}
\newcommand{\Anuo}{{A^{\nu}}}
\newcommand{\Vnuu}{{V_{\nu}}}
\newcommand{\Vnuo}{{V^{\nu}}}
\newcommand{\Fmnu}{{F_{\mu\nu}}}
\newcommand{\Fmno}{{F^{\mu\nu}}}

\newcommand{\sgn}{\mathop{\mathrm{sgn}}}

\newcommand{\abcd}{{\alpha\beta\gamma\delta}}


\newcommand{\bsigma}{\mbox{\boldmath $\sigma$}}
\newcommand{\beps}{\mbox{\boldmath $\varepsilon$}}
\newcommand{\btau}{\mbox{\boldmath $\tau$}}
\newcommand{\brho}{\mbox{\boldmath $\rho$}}
\newcommand{\bpipi}{\mbox{\boldmath $\pi\pi$}} 
\newcommand{\bss}{\bsigma\!\cdot\!\bsigma}
\newcommand{\btt}{\btau\!\cdot\!\btau}
\newcommand{\bnabla}{\mbox{\boldmath $\nabla$}}
\newcommand{\bphi}{\mbox{\boldmath $\tau$}}
\newcommand{\bvarphi}{\mbox{\boldmath $\rho$}}
\newcommand{\bE}{\mbox{\boldmath $E$}}
\newcommand{\bDelta}{\mbox{\boldmath $\Delta$}}
\newcommand{\bGamma}{\mbox{\boldmath $\Gamma$}}
\newcommand{\bpsi}{\mbox{\boldmath $\psi$}}
\newcommand{\bPsi}{\mbox{\boldmath $\Psi$}}
\newcommand{\bPhi}{\mbox{\boldmath $\Phi$}}
\newcommand{\bnab}{\mbox{\boldmath $\nabla$}}
\newcommand{\bpi}{\mbox{\boldmath $\pi$}}
\newcommand{\btheta}{\mbox{\boldmath $\theta$}}
\newcommand{\bkappa}{\mbox{\boldmath $\kappa$}}
\newcommand{\bgamma}{\mbox{\boldmath $\gamma$}}

\newcommand{\bp}{\mbox{\boldmath $p$}}
\newcommand{\ba}{\mbox{\boldmath $a$}}
\newcommand{\bq}{\mbox{\boldmath $q$}}
\newcommand{\br}{\mbox{\boldmath $r$}}
\newcommand{\bs}{\mbox{\boldmath $s$}}
\newcommand{\bk}{\mbox{\boldmath $k$}}
\newcommand{\bl}{\mbox{\boldmath $l$}}
\newcommand{\bb}{\mbox{\boldmath $b$}}
\newcommand{\be}{\mbox{\boldmath $e$}}
\newcommand{\bP}{\mbox{\boldmath $P$}}
\newcommand{\bV}{\mbox{\boldmath $V$}}
\newcommand{\bI}{\mbox{\boldmath $I$}}
\newcommand{\bJ}{\mbox{\boldmath $J$}}

\newcommand{\bT}{\mbox{\boldmath $\mathcal{T}$}}
\newcommand{\fph}{${\cal F}$}
\newcommand{\aph}{${\cal A}$}
\newcommand{\dph}{${\cal D}$}
\newcommand{\fpi}{f_\pi}
\newcommand{\mpi}{m_\pi}
\newcommand{\Tr}{{\mbox{\rm Tr}}}
\def\Qb{\overline{Q}}
\newcommand{\delu}{\partial_{\mu}}
\newcommand{\delo}{\partial^{\mu}}
\newcommand{\up}{\!\uparrow}
\newcommand{\upup}{\uparrow\uparrow}
\newcommand{\updo}{\uparrow\downarrow}
\newcommand{\uu}{$\uparrow\uparrow$}
\newcommand{\ud}{$\uparrow\downarrow$}
\newcommand{\auu}{$a^{\uparrow\uparrow}$}
\newcommand{\aud}{$a^{\uparrow\downarrow}$}
\newcommand{\pu}{p\!\uparrow}
\newcommand{\qp}{quasiparticle}
\newcommand{\sa}{scattering amplitude}
\newcommand{\ph}{particle-hole}
\newcommand{\qcd}{{\it QCD}}
\newcommand{\integ}{\int\!d}
\newcommand{\ie}{{\sl i.e.~}}
\newcommand{\etal}{{\sl et al.~}}
\newcommand{\etc}{{\sl etc.~}}
\newcommand{\rhs}{{\sl rhs~}}
\newcommand{\lhs}{{\sl lhs~}}
\newcommand{\eg}{{\sl e.g.~}}
\newcommand{\ef}{\epsilon_F}
\newcommand{\sigt}{d^2\sigma/d\Omega dE}
\newcommand{\sige}{{d^2\sigma\over d\Omega dE}}
\newcommand{\rpaeq}{\beq
\left ( \begin{array}{cc}
A&B\\
-B^*&-A^*\end{array}\right )
\left ( \begin{array}{c}
X^{(\kappa})\\Y^{(\kappa)}\end{array}\right )=E_\kappa
\left ( \begin{array}{c}
X^{(\kappa})\\Y^{(\kappa)}\end{array}\right )
\eeq}

\newcommand{\ket}[1]{{#1} \rangle}
\newcommand{\bra}[1]{\langle {#1} }

\newcommand{\Bigket}[1]{{#1} \Big\rangle}
\newcommand{\Bigbra}[1]{\Big\langle {#1} }

\newcommand{\ave}[1]{\langle {#1} \rangle}
\newcommand{\Bigave}[1]{\left\langle {#1} \right\rangle}
\newcommand{\half}{{1\over 2}}

\newcommand{\singlespace}{
    \renewcommand{\baselinestretch}{1}\large\normalsize}
\newcommand{\doublespace}{
    \renewcommand{\baselinestretch}{1.6}\large\normalsize}
\newcommand{\bftau}{\mbox{\boldmath $\tau$}}
\newcommand{\bfalpha}{\mbox{\boldmath $\alpha$}}
\newcommand{\bfgamma}{\mbox{\boldmath $\gamma$}}
\newcommand{\bfxi}{\mbox{\boldmath $\xi$}}
\newcommand{\bfbeta}{\mbox{\boldmath $\beta$}}
\newcommand{\bfeta}{\mbox{\boldmath $\eta$}}
\newcommand{\bfpi}{\mbox{\boldmath $\pi$}}
\newcommand{\bfphi}{\mbox{\boldmath $\phi$}}
\newcommand{\bfR}{\mbox{\boldmath ${\cal R}$}}
\newcommand{\bfL}{\mbox{\boldmath ${\cal L}$}}
\newcommand{\bfM}{\mbox{\boldmath ${\cal M}$}}
\def\dblint{\mathop{\rlap{\hbox{$\displaystyle\!\int\!\!\!\!\!\int$}}
    \hbox{$\bigcirc$}}}
\def\ut#1{$\underline{\smash{\vphantom{y}\hbox{#1}}}$}

\def\UNITY{{\bf 1\! |}}
\def\Pom{{\bf I\!P}}
\def\lsim{\mathrel{\rlap{\lower4pt\hbox{\hskip1pt$\sim$}}
    \raise1pt\hbox{$<$}}}         
\def\gsim{\mathrel{\rlap{\lower4pt\hbox{\hskip1pt$\sim$}}
    \raise1pt\hbox{$>$}}}         

\newcommand\scalemath[2]{\scalebox{#1}{\mbox{\ensuremath{\displaystyle #2}}}}

\title{Exclusive $\eta_c$ production by $\gamma^{*} \gamma$ interactions \\ in electron-ion collisions}

\author{Izabela Babiarz}
\email{izabela.babiarz@ifj.edu.pl.pl}
\affiliation{Institute of Nuclear Physics Polish Academy of Sciences, 
ul. Radzikowskiego 152, PL-31-342 Krak{\'o}w, Poland}

\author{Victor P.  Goncalves}
\email{barros@ufpel.edu.br }
\affiliation{Institut f\"ur Theoretische Physik, Westf\"alische Wilhelms-Universit\"at M\"unster,
Wilhelm-Klemm-Stra\ss e 9, D-48149 M\"unster, Germany}
\affiliation{Physics and Mathematics Institute, Federal University of Pelotas, \\
  Postal Code 354,  96010-900, Pelotas, RS, Brazil}

\author{Wolfgang Sch\"afer}
\email{Wolfgang.Schafer@ifj.edu.pl} 
\affiliation{Institute of Nuclear
Physics Polish Academy of Sciences, ul. Radzikowskiego 152, PL-31-342 
Krak{\'o}w, Poland}

\author{Antoni Szczurek}
\email{antoni.szczurek@ifj.edu.pl}
\affiliation{College of Mathematics and Natural Sciences,
University of Rzesz\'ow, ul. Pigonia 1, PL-35-310 Rzesz\'ow, Poland\vspace{5mm}}


\begin{abstract}
One of the main goals of future electron-ion colliders is to improve our understanding of the structure of hadrons. In this letter, we study the exclusive $\eta_c$ production by $\gamma^{*} \gamma$ interactions in $eA$ collisions and demonstrate that future experimental analysis of this process can be used to improve the description of the $\eta_c$ transition form factor. The rapidity, transverse momentum and photon virtuality distributions are estimated considering the energy and target configurations expected to be present at the EIC, EicC and LHeC and assuming different predictions for the light-front wave function of the $\eta_c$ meson. Our results indicate that the electron-ion colliders can be considered an alternative to providing supplementary data to those  obtained in $e^- e^+$ colliders.
\end{abstract}

\maketitle

\section{Introduction}

The improvement of our understanding of the quantum 3D imaging of the partons inside the protons and nuclei, encoded in the quantum phase space Wigner distributions, which include information on both generalised parton distributions (GPDs) and transverse momentum parton distributions (TMDs), is one of the main goals of the future electron-ion colliders at BNL (EIC) \cite{Boer:2011fh,*Accardi:2012qut,*Aschenauer:2017jsk,*AbdulKhalek:2021gbh,*Burkert:2022hjz,*Abir:2023fpo}, CERN (LHeC) \cite{LHeCStudyGroup:2012zhm} and China (EicC) \cite{EicC}. A tomography picture of the hadrons is expected to be revealed in deep inelastic electron-hadron scattering by measurements of exclusive processes, wherein the hadron remains intact after scattering by the lepton probe. One of the promising final states is the exclusive production of heavy vector mesons ($J/\Psi, \, \Psi(2S), \, \Upsilon, ...$), which occurs by the exchange of a color singlet object with vacuum quantum numbers ($C = P = +1$, with $C$ being the charge conjugation and $P$ the parity) \cite{Barone:2002cv}. In the color dipole formalism, \cite{Nikolaev:1992si,*Kopeliovich:1993gk},
the scattering amplitude for this process is described in terms of the light-front wave functions (LFWFs) for the photon and the vector meson and the dipole - hadron cross-section, which is determined by the theory of  strong interactions. As a consequence, the associated cross-section is strongly sensitive to the underlying QCD dynamics and the description of the vector meson structure. Such aspects have motivated extensive phenomenology in the last years (See, e.g. Refs. \cite{Armesto:2014sma,Cisek:2014ala,run2,Chen:2016dlk,ArroyoGarcia:2019cfl,Goncalves:2020ywm,Shi:2021taf,Mantysaari:2022kdm}), with the results indicating that the future electron-ion colliders, characterised by high center - of - mass  energies and/or large luminosities, will be able to constrain the theoretical description of this process.

\begin{figure}[t]
{\includegraphics[width=0.55\textwidth]{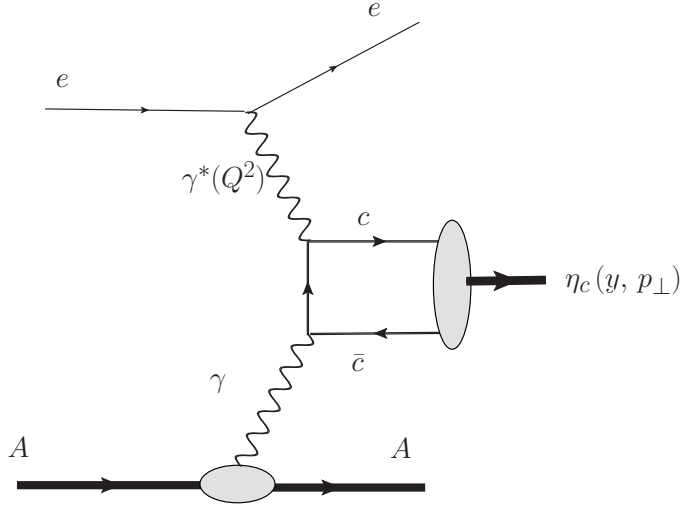}}  
\caption{Exclusive $\eta_c$ production by $\gamma^* \gamma$ interactions in electron - ion collisions.}
\label{fig:etac_eic}
\end{figure}

An important open question is if these future colliders could also be used to improve our understanding of the structure of pseudoscalar mesons. As this final state is characterised by a positive $C$ parity, in order to be produced in an exclusive reaction induced by a photon, the object exchanged in the $t$ - channel should have negative $C$ parity as, e.g. a photon or an Odderon.  In perturbative QCD, the Odderon is described by a compound state of three Reggeized gluons \cite{Barone:2002cv}, with evolution given by the Bartels-Kwiecinski-Praszalowicz (BKP) equation \cite{Bartels:1980pe,*Kwiecinski:1980wb}. During the last years, the existence of the Odderon has been a theme of intense debate in the literature (for a review see, e.g., Ref. \cite{Ewerz:2003xi}), with several studies \cite{Kilian:1997ew,nac,nac_odd,Czyzewski:1996bv,Engel:1997cga,bbcv,vic_odderon1,vic_odderon2,vicbru_etac,Dumitru:2019qec} indicating that future experimental analysis of the exclusive $\eta_c$ production in $ep$ and ultraperipheral collisions could be useful to improve our understanding of Odderon. Such studies also indicate that the contribution for the $\eta_c$ production associated with photon-photon interactions is  similar for a proton target and dominates when an ion is present, which is directly associated with the $Z^2$ enhancement present in the nuclear photon flux ($Z$ is the nuclear charge). This aspect strongly motivates the analysis of the exclusive $\eta_c$ production in $eA$ collisions, which can occur through the interaction between a virtual photon emitted by the electron and a (quasi-) real photon emitted by the nucleus, as represented in Fig. \ref{fig:etac_eic}.
As we will demonstrate below, the associated cross-section can be expressed in terms of the $\eta_c$ transition form factor $F(Q_1^2= Q^2, Q_2^2 = 0)$, which is determined by the $\eta_c$ LFWF  and, therefore, can be used to provide  
valuable information on the structure of pseudoscalar mesons. In this exploratory study, we will use the formalism developed in Refs. \cite{Babiarz:2019sfa,Babiarz:2019mag,Babiarz:2020jkh,Babiarz:2022xxm,Babiarz:2023ebe}, where the transition form factor was estimated  
considering light-front wave functions derived from different
potential models and the predictions have been compared with the 
data from the BaBar collaboration
\cite{BaBar:2010siw} extracted from the $e^+ e^- \to e^+ e^- \eta_c$ reaction. Our results indicate that the future electron-ion colliders can provide supplementary data to those obtained in $e^+ e^-$ collisions.

This letter is organised in the following order. In the next Section, we will present a brief review of the formalism needed to describe the exclusive $\eta_c$ production in $eA$ collisions, with particular emphasis on the description of the transition form factor developed in Ref. \cite{Babiarz:2019sfa}. 
In Section \ref{sec:res}, we present our predictions for the rapidity, transverse momentum and $Q^2$ distributions considering the expected energy and target configurations for the EIC, LHeC and EicC and assuming distinct potential models to estimate the transition form factor. Finally, in Section \ref{sec:conc}, we summarise our main results and conclusions.
\section{$\gamma^* \gamma$ cross-section and transition form factor}
The total cross-section for the exclusive $\eta_c$ production in electron-ion collisions can be factorised as follows
\begin{eqnarray}
    \sigma(e A \rightarrow e \eta_c A) = \int d\omega_e dQ^2 \frac{d^2N_e}{d\omega_e dQ^2} \, \sigma(\gamma^* A \rightarrow \eta_c A)\, ,
\end{eqnarray}
with the photon flux for an electron being given by \cite{Budnev:1975poe} 
\begin{eqnarray}
\frac{d^2N_e}{d\omega_e dQ^2} =    \frac{\alpha_{em}}{\pi \omega_e Q^2} \left[\left(1 - \frac{\omega_e}{E_e}\right)\left(1 - \frac{Q^2_{min}}{Q^2}\right) + \frac{\omega_e^2}{2E_e^2}\right]\,,
\end{eqnarray}
where $\omega_e$ is the energy of the photon emitted by the electron with energy $E_e$, and $Q^2$ is its virtuality. Moreover, one has that $Q_{min}^2 = m_e^2\omega_e^2/[E_e(E_e - \omega_e)]$ and $Q_{max}^2 = 4E_e(E_e - \omega_e)$, which is constrained by the maximum of the electron energy loss. Assuming that the $\eta_c$ production will be dominated by the $\gamma^* \gamma \rightarrow \eta_c $ subprocess, one has that the equivalent photon approximation implies that
\begin{eqnarray}
    \sigma(\gamma^* A \rightarrow \eta_c A) = \int d\omega_A \, \frac{dN}{d\omega_A} \, \sigma_{\rm TT} 
    (\gamma^* \gamma \rightarrow \eta_c; W_{\gamma \gamma},Q^2,0)\, ,
\end{eqnarray}
where the photon flux for the nucleus for a photon with energy $\omega_A$ is given by 
\cite{Bertulani:1987tz,*Bertulani:2005ru,*Goncalves:2005sn,*Baltz:2007kq,*Contreras:2015dqa}
\begin{eqnarray}
    \frac{dN}{d\omega_A} = \frac{2 Z^2 \alpha_{em}}{\pi \omega_A} \left[\xi K_0(\xi)K_1(\xi) - \frac{\xi^2}{2}(K_1^2(\xi) - K_0^2 (\xi))\right]\, ,
\end{eqnarray}
with $\xi = R_A \omega_A/\gamma_L$, where $R_A = r_0 A^{1/3}$, with $r_0 = 1.1\,{\rm fm}$, is the nuclear radius, $\gamma_L$ is the Lorentz factor and $K_0$ and $K_1$
are the modified Bessel functions. In addition, one has that photon-photon center of mass energy $W_{\gamma \gamma}$ will be $W_{\gamma \gamma} = \sqrt{4 \omega_e \omega_A - p^2_{\perp}}$, where $p_{\perp}$ is the transverse momentum of the meson in the final state.
Moreover, in the $eA$ cm-frame, the photon energies $\omega_i$ can be written in terms of the rapidity
$y$ of the final state as follows 
\begin{eqnarray}
    \omega_e = \frac{\sqrt{M_{\eta_c}^2+p_{\perp}^2}}{2} e^{+y}   \,\,\,\,\,\, \mbox{and} \,\,\,\,\,\,  \omega_A = \frac{\sqrt{M_{\eta_c}^2 + p_{\perp}^2}}{2} e^{-y} \,,
    \label{eq:omegas}
\end{eqnarray}
with the meson transverse momentum being given by
\begin{eqnarray}
    p_{\perp}^2 = \Big( 1-\frac{\omega_e}{E_e} \Big) Q^2\, .
\end{eqnarray}
The main input to calculate the exclusive $\eta_c$ production in electron-ion collisions is the cross-section for the  $\gamma^* \gamma \rightarrow \eta_c $ subprocess. As discussed in detail in Refs.~\cite{Budnev:1975poe, Poppe1986}, the cross-section for the general case of the interaction between two virtual photons can be expressed as
\begin{eqnarray}
    \sigma_{\rm TT}(W_{\gamma \gamma},Q_1^2,Q_2^2)
    = {1 \over 4 \sqrt{X}} \,  
  \frac {M_{\eta_c} \Gamma_{\rm tot}}{(W^2_{\gamma \gamma} - M_{\eta_c}^2)^2 + M_{\eta_c}^2 \Gamma^2_{\rm tot} } \, {\mathcal M}^*(++) {\mathcal M}(++) \, , 
\end{eqnarray}
where $ M_{\eta_c}$ is the mass of the $\eta_c$ meson and $X = (q_1\cdot q_2)^2 - q_1^2 q_2^2$, with the photon virtualities being defined by $Q_i^2 = - q_i^2$. Moreover, the helicity amplitude ${\cal M} (\lambda_1,\lambda_2)$  can be written as
\begin{eqnarray}
    {\cal M} (\lambda_1,\lambda_2) = e^1_\mu(\lambda_1) e^2_\nu(\lambda_2) {\mathcal M}^{\mu \nu} \, ,
\end{eqnarray}
with the polarization vectors $e^{1,2}_\mu(\lambda_{1,2})$ defined in $\gamma^{*}\gamma^{*}$ center of mass frame, as in \cite{Babiarz:2022xxm}.
The transition form factor can be obtained from  the covariant amplitude as follows \cite{Poppe1986}:
\begin{eqnarray}
{\cal M}_{\mu \nu}(\gamma^*(q_1) \gamma^*(q_2) \to \eta_c) 
= 4 \pi \alpha_{\rm em} \, (-i) \varepsilon_{\mu \nu \alpha \beta} q_1^\alpha q_2^\beta \, F(Q_1^2, Q_2^2) \, .
\end{eqnarray}
In the $\gamma^* \gamma^{*}$ c.m. frame, with $\hat x_\mu, \hat y_\mu$ being spacelike directions orthogonal to the collision axis, we can write
\begin{eqnarray}
    {\mathcal M}_{\mu \nu} = i \, 4 \pi \alpha_{\rm em} \, \sqrt{X} \,\Big(\hat x_\mu \hat y_\nu - \hat y_\mu \hat x_\nu \Big) \, F (Q_1^2,Q_2^2) \, . 
\end{eqnarray}
In the limit of interest in this letter, where $Q_1^2 = Q^2$, $Q_2^2 = 0$, one has that $\sqrt{X} = q_1 \cdot q_2 = (M_{\eta_c}^2 + Q^2)/2$ and the  $\gamma^* \gamma$ cross section is expressed by \cite{Babiarz:2019sfa}:
\begin{eqnarray}
    \sigma_{\rm TT} (W_{\gamma \gamma},Q^2,0) = 2 \pi^2 \alpha^2_{\rm em} \,  \frac {M_{\eta_c} \Gamma_{\rm tot}}{(W^2_{\gamma \gamma} - M_{\eta_c}^2)^2 + M_{\eta_c}^2 \Gamma^2_{\rm tot} } \, 
    (M_{\eta_c}^2 + Q^2) \, F^2(Q^2,0) \, .
\end{eqnarray}
Using the relation between the $\gamma \gamma$ decay width and the form factor at the on-shell point, 
\begin{eqnarray}
 \Gamma_{\gamma \gamma} = \frac{\pi \alpha^2_{\rm em} M^3_{\eta_c}}{4} \, F^2(0,0)  \, ,    
\end{eqnarray}
it is possible to rewrite the previous equation in the form
\begin{eqnarray}
    \sigma_{\mathrm{TT}} (W_{\gamma \gamma},Q^2,0) &=& 8 \pi \frac {\Gamma_{\gamma \gamma} \Gamma_{\rm tot}}{(W^2_{\gamma \gamma} - M_{\eta_c}^2)^2 + M_{\eta_c}^2 \Gamma^2_{\rm tot} } \, \Big(1 + \frac{Q^2}{M_{\eta_c}^2}\Big) \Big( \frac{F(Q^2,0)}{F(0,0)} \Big)^2 \, \nonumber \\
    &\approx& 8 \pi^2 \, \delta(W^2_{\gamma \gamma} - M_{\eta_c}^2) \, \frac{\Gamma_{\gamma \gamma}}{M_{\eta_c}} \, \Big(1 + \frac{Q^2}{M_{\eta_c}^2}\Big) \Big( \frac{F(Q^2,0)}{F(0,0)} \Big)^2 \, ,
    \label{eq:sig_TT}
\end{eqnarray}
Here, in the last line, we have adopted the narrow-width approximation,  
which  reduces to the well-known Low formula \cite{Low:1960wv}  for $Q^2 = 0$, and implies that the cross-section for the $\gamma^* A \rightarrow \eta_c A$ reaction reads:
\begin{eqnarray}
\displaystyle
\sigma(\gamma^* A \to \eta_c A) = \frac{dN}{d\omega_A}{\Bigg|}_{\omega_A = (M_{\eta_c}^2 + p^2_{\perp})/(4 \omega_e)}8\pi^2 \frac{1}{4\omega_e} \frac{\Gamma_{\gamma\gamma}}{M_{\eta_c}} \Big( 1+\frac{Q^2}{M_{\eta_c}^2} \Big) \Big( \frac{F(Q^2,0)}{F(0,0)} \Big)^2 \, .
\label{eq:sigma_gamma_A}
\end{eqnarray}
\begin{figure}[t]
{\includegraphics[width=0.5\textwidth]{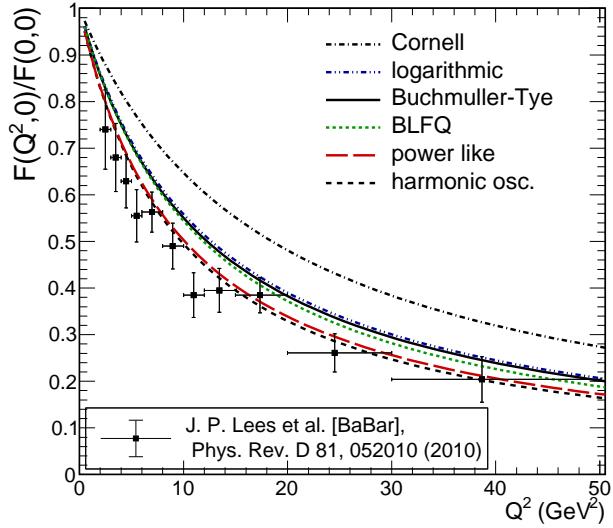}}  
\caption{Dependence of the normalised transition form factor, $F(Q^2,0)/F(0,0)$, on the photon virtuality $Q^2$ predicted by the different potential models considered in Ref. \cite{Babiarz:2019sfa}. The prediction of the BLFQ approach \cite{Li:2021ejv,Li:2017mlw} and the BaBar data \cite{BaBar:2010siw} are also presented  for comparison. }
\label{fig:efe}
\end{figure}
This expression puts into evidence that the exclusive $\eta_c$ cross-section in $eA$ collisions will be determined by the $\eta_c$ transition form factor for one virtual photon $F(Q^2,0)$. 
Here we adopt the light-front approach, where the transition form factor can be expressed through the LFWF depending on the LF momentum fraction $z$ and transverse momentum $\vec k_\perp$ of quarks in the bound state. For the 
spinless meson, we have for the $c \bar c$ Fock-state wave function 
\begin{eqnarray}
    \Psi_{\lambda \bar \lambda}(z, \vec k_\perp) = e^{i m \phi} \, \tilde \psi_{\lambda \bar \lambda} (z,k_\perp) \, ,
\end{eqnarray}
where $\vec k_\perp = k_\perp(\cos \phi, \sin \phi)$, and $m = |\lambda + \bar \lambda|$. Below we denote the light-front helicities of quark and antiquark, $ \lambda, \bar \lambda$ by $\uparrow, \downarrow$ for their values $\lambda, \bar \lambda = \pm 1/2$.

As demonstrated in Ref.~\cite{Babiarz:2019sfa,Babiarz:2021ung}, in the limit where $Q^2_2 \to 0$, transition form factor is:
\begin{multline}
    F(Q^2, 0) =  e_c^2 \sqrt{N_c}\, 4
    \int \frac{dz d^2 \vec{k}_{\perp}}{\sqrt{z(1-z)}16\pi^3}
    \Bigg\{ 
    \frac{1}{\vec{k}_{\perp}\,^2 + \mu^2} \tilde{\psi}_{\uparrow\downarrow}(z, k_{\perp})
    \\
    + \frac{\vec{k}_{\perp}\,^2}{[\vec{k}_{\perp}\,^2 + \mu^2]^2}
    \Big( \tilde{\psi}_{\uparrow\downarrow}(z, k_{\perp}) + \frac{m_c}{k_{\perp}} \tilde{\psi}_{\uparrow\uparrow}(z, k_{\perp})  \Big) \Bigg\} \, ,   
    \label{eq:FF_LF}
\end{multline}
with $\mu^2 = z(1-z) Q^2 + m_c^2$. 

In the Melosh spin-rotation formulation used in Ref. \cite{Babiarz:2019sfa}, the helicity components $\tilde{\psi}_{\uparrow\downarrow}(z, k_{\perp}),\tilde{\psi}_{\uparrow\uparrow}(z, k_{\perp})
$,  are related to the same radial wave function~$\psi(z,k_\perp)$~as:
\begin{eqnarray}
    \tilde{\psi}_{\uparrow\downarrow}(z, k_{\perp}) \rightarrow \frac{m_c}{\sqrt{z(1-z)}}\, \psi(z,k_{\perp})\,, \quad {\rm and}\quad \tilde{\psi}_{\uparrow\uparrow}(z, k_{\perp}) \rightarrow \frac{-|\vec{k}_{\perp}|}{\sqrt{z(1-z)}}\, \psi(z,k_{\perp}) \, ,
\end{eqnarray}
so that there appears a cancellation of the terms in the round brackets in Eq.~(\ref{eq:FF_LF}) \cite{Babiarz:2019sfa}. This cancellation is not guaranteed in other approaches, such as the BLFQ (Basis Light-Front Quantization) approach of \cite{Li:2017mlw, Li:2021ejv}.

In our numerical calculations, we use the analytical form of the harmonic oscillator wave function, as well as the radial WFs for a variety of interquark potentials from Ref.\cite{Cepila:2019skb}. 
In Ref. \cite{Babiarz:2019sfa}, the LFWFs were discussed in detail, and the relevant expressions were explicitly presented.

In addition to these, we also adopt the helicity WFs in the BLFQ approach using the tables of Ref.\cite{Li:2017MenData}. It turns out that also, for these LFWFs, the cancellation of the 
terms in round brackets in Eq.~(\ref{eq:FF_LF}) is nearly exact.

In Fig. \ref{fig:efe}, we present a comparison between the predictions derived using these distinct models. In addition to the results already shown in \cite{Babiarz:2019sfa}, we also present the prediction derived in the BLFQ approach of Ref. \cite{Li:2017mlw}. Also shown are the BaBar data \cite{BaBar:2010siw} for the normalized from factor extracted in the analysis of the $e^+ e^- \to e^+ e^- \eta_c$ reaction.

As already observed in Ref. \cite{Babiarz:2019sfa}, the oscillator and power-law potentials give the best description of the BaBar data. However, it is important to emphasise that a better description of the data can also be obtained if a smaller value for the charm mass is used in the other potentials. A very good description is also obtained using the BLFQ approach \cite{Li:2017mlw}. Surely, more data, with a larger precision, will allow us to improve our understanding of the meson structure.
Regarding the normalization of the cross-section of Eq.~(\ref{eq:sigma_gamma_A}), which is sensitive to the $\gamma \gamma$ decay width of the $\eta_c$, it is worth emphasizing that the value of the latter is still under debate, see for example the recent discussion in \cite{Colquhoun:2023zbc}.
In our calculations, we use the PDG average
\cite{PDG}: $ \Gamma_{\gamma \gamma} = (5.4\pm 0.4)\,\rm{keV}  \,.$
\section{Numerical results for electron-ion collisions}
\label{sec:res}
\begin{figure}[hpt!]
    \centering
    \includegraphics[width =\textwidth]{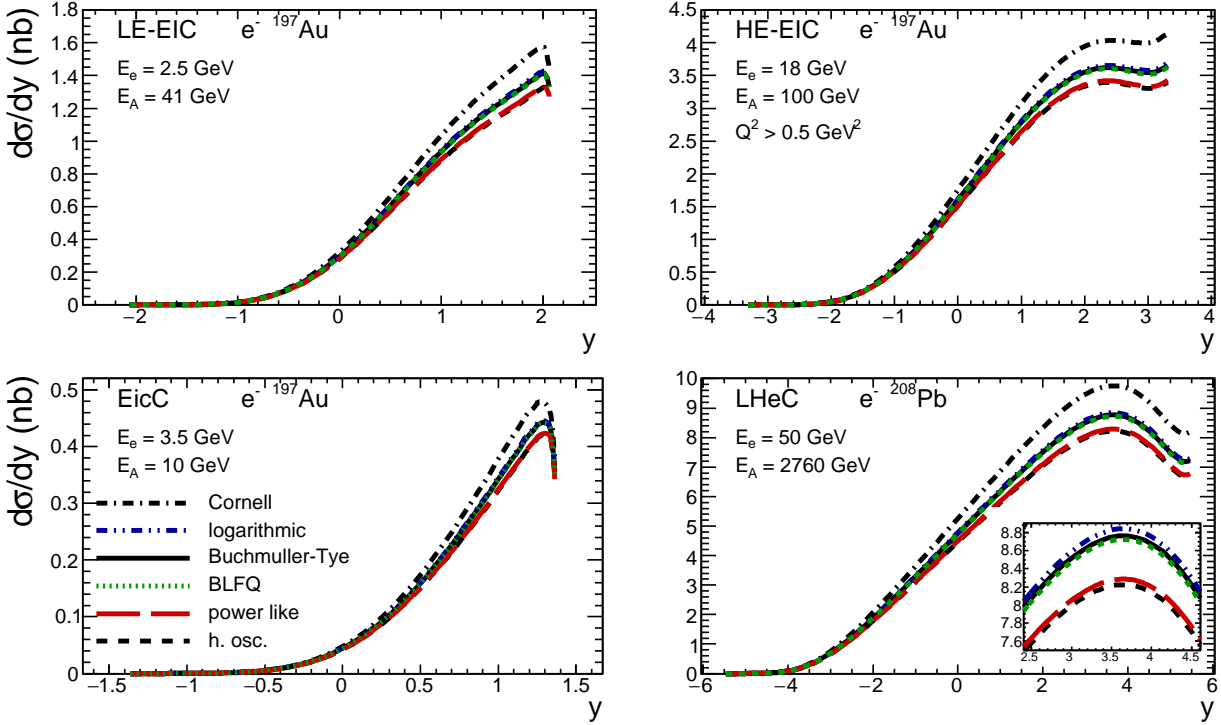}
    \caption{Distributions of the eA cm-frame rapidity for the exclusive $\eta_c$ production in electron-ion collisions at the LE - EIC, HE - EIC, EicC and LHeC derived considering distinct models for the transition form factor. }
    \label{fig:dsig_dy}
\end{figure}
\begin{figure}[h!]
    \centering
    \includegraphics[width = \textwidth]{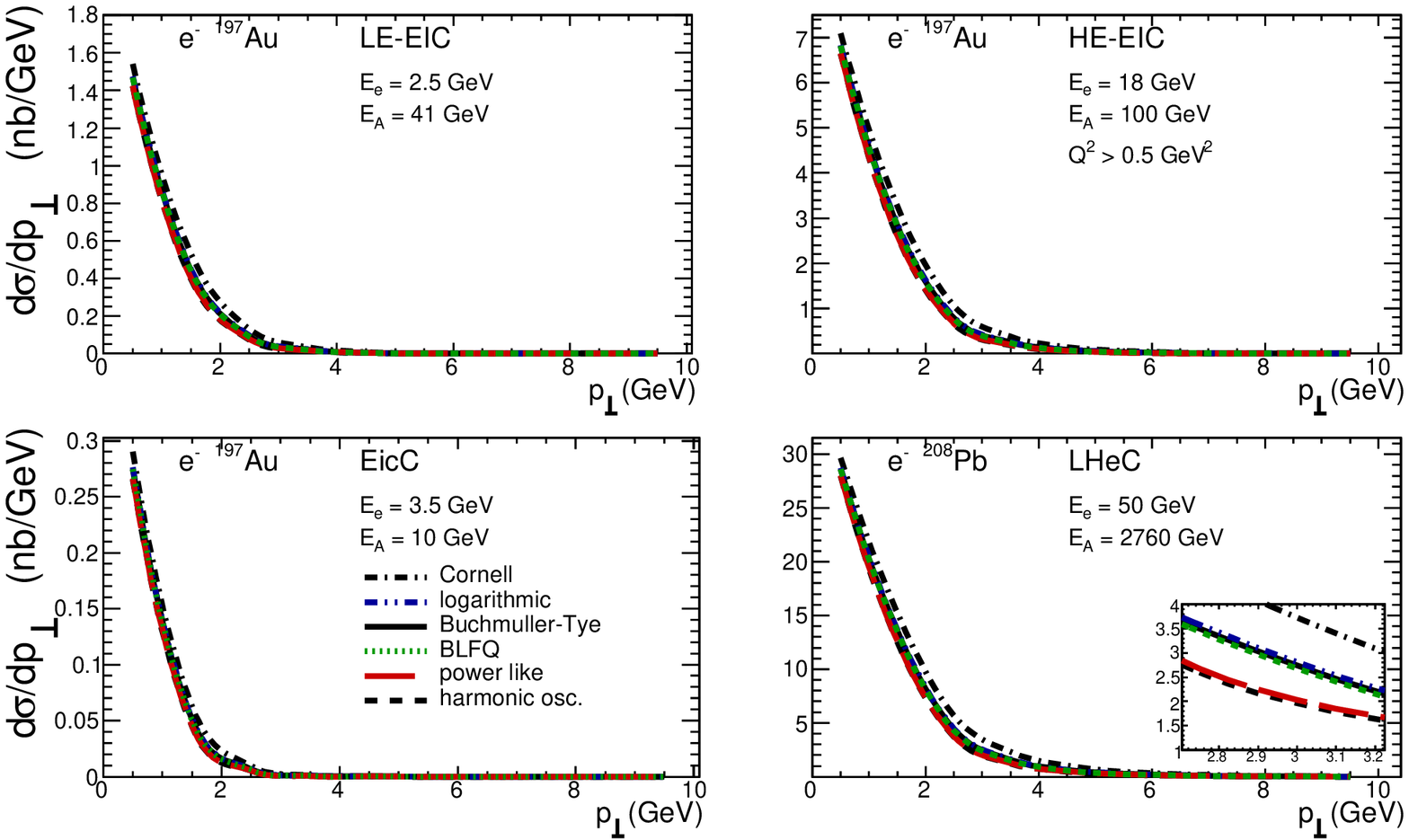}
    \caption{Transverse momentum distributions for the exclusive $\eta_c$ production in electron-ion collisions at the LE - EIC, HE - EIC, EicC and LHeC derived considering distinct models for the transition form factor. }
    \label{fig:dsig_dpt}
\end{figure}
In what follows, we will estimate the rapidity, transverse momentum and $Q^2$ distributions considering the energy and target configurations expected in the future electron-ion colliders at the BNL, CERN and in China. These distributions will be estimated considering the predictions for the $\eta_c$ transition form factor obtained by solving the Schr\"odinger equation for distinct models of the $c\bar{c}$ potential used in Ref. \cite{Babiarz:2019sfa}. In addition, we will also present the predictions derived using the results obtained in Ref. \cite{Li:2017mlw}.
As the future electron-ion collider at BNL,  the electron beam with an energy up to 18 GeV will be set to collide  with a heavy ion with energies up to 100 GeV \cite{*Boer:2011fh,*Accardi:2012qut,*Aschenauer:2017jsk,*AbdulKhalek:2021gbh,*Burkert:2022hjz,*Abir:2023fpo}, reaching  luminosities in the $10^{33} - 10^{34}$ cm$^{-2}$s$^{-1}$ range, in our analysis, we will assume in the following two distinct benchmarks for the electron and $Au$ - ion energies: (a) $(E_e,\, E_{Au}) = (2.5,\, 41)$ GeV and (b) $(E_e,\, E_{Au}) = (18,\, 100)$ GeV. These configurations will be denoted hereafter LE - EIC and HE - EIC, respectively. Moreover, we will also estimate the distributions for the EicC \cite{EicC} ($E_e = 3.5$ GeV, $E_{Au} = 10$ GeV and ${\cal{L}} = 10^{33}$ cm$^{-2}$s$^{-1})$ and for the  LHeC \cite{LHeCStudyGroup:2012zhm} ($E_e = 50$ GeV, $E_{Pb} = 2760$ GeV and   ${\cal{L}} = 10^{32}$ cm$^{-2}$s$^{-1}$).

In Fig. \ref{fig:dsig_dy}, we present our predictions for the rapidity distributions of the exclusive $\eta_c$ production in electron-ion collisions at the LE - EIC, HE - EIC, EicC and LHeC, estimated integrating over the photon virtuality in the range $Q^2 > 0.5$ GeV$^2$. One has that the distribution increases with rapidity, with the maximum occurring at larger values when the center-of-mass energies are increased. 
As expected from Fig. \ref{fig:efe}, one has that the harmonic oscillator and power-like predictions are similar and smaller than those derived assuming the other models for the transition form factor. The difference between the predictions increases with the rapidity and the center-of-mass energy.  

The predictions for the transverse momentum distribution are presented in Fig. \ref{fig:dsig_dpt}.
One has that the cross-section increases with the center-of-mass energy and decreases with $p_{\perp}$,  being strongly peaked for $p_{\perp} \rightarrow 0$ and with the difference between the predictions increasing for larger values of the transverse momentum. 
\begin{figure}[h!]
    \centering
    \includegraphics[width =\textwidth]{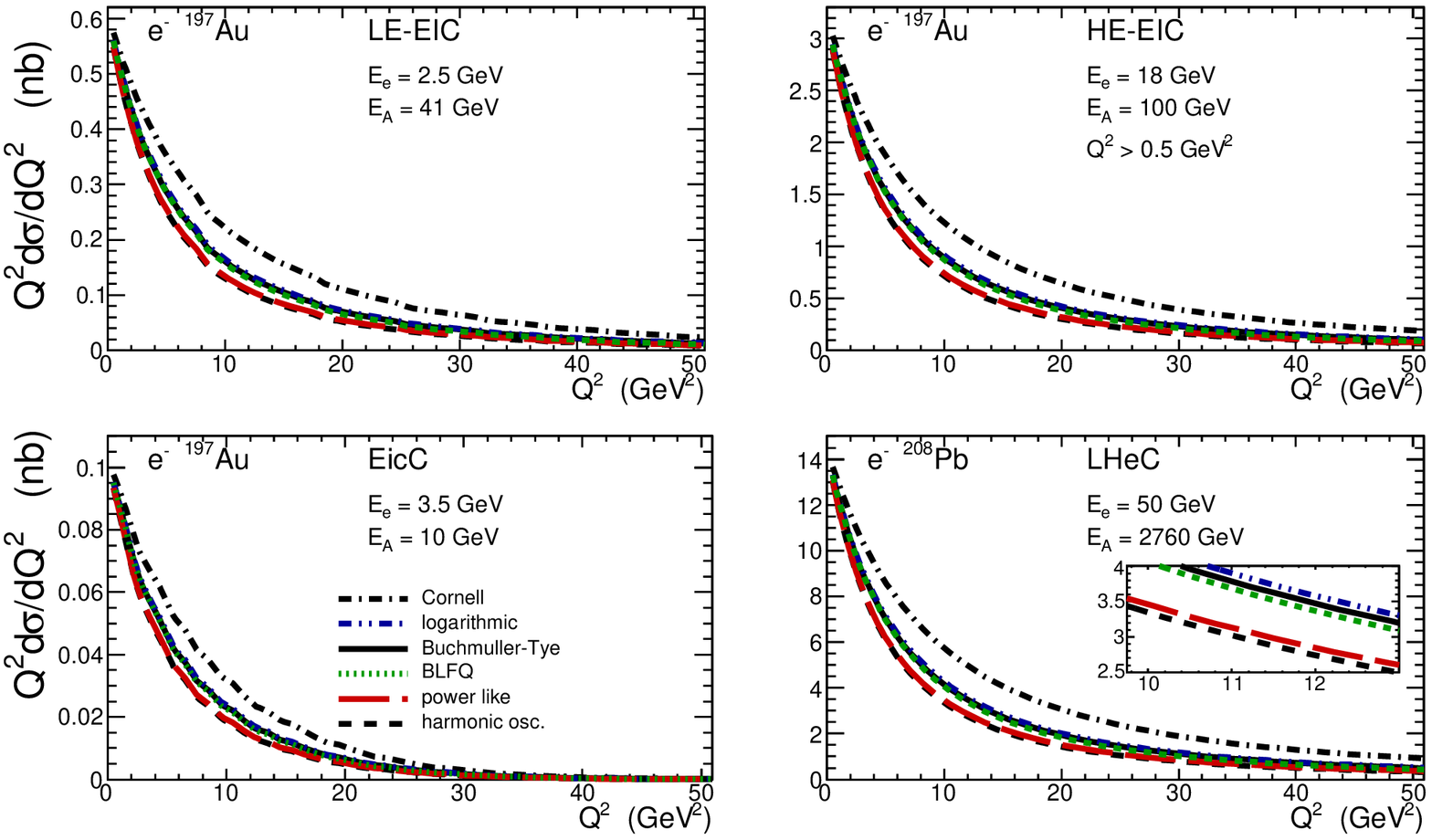}
    \caption{Differential $Q^2$ distributions for the exclusive $\eta_c$ production in electron-ion collisions at the LE - EIC, HE - EIC, EicC and LHeC derived considering distinct models for the transition form factor. }
    \label{fig:Q2dsig_dQ2}
\end{figure}
In Fig. \ref{fig:Q2dsig_dQ2} we present our predictions for the dependence on $Q^2$ of the $Q^2 d\sigma/dQ^2$ distribution, which can be measured by tagging the electron in the final state. For the Cornell model we obtain generally larger values. The results fore the harmonic oscillator and power-like predictions are rather similar to each other,
while the BLFQ, Buchm{\"u}ller-Tye and logarithmic models do not differ much amongst each other. As in the previous distributions, the difference between the predictions increases with the center-of-mass energies. Such results indicate that the analysis of the  distributions considered in this letter can be useful to constrain the transition form factors.

A final comment is in order. The results derived in this paper indicate that the cross sections for the future electron-ion colliders are of the order of 0.1 -- 60 nb. Considering these values and the expected luminosities, we predict that the number of events per year will be ${\cal{O}}(10^6)$ and ${\cal{O}}(10^7)$ for the EicC and LHeC, respectively. In contrast, we predict that the number of events per year will be one order of magnitude larger for the EIC. Such larger numbers imply that a future experimental study of the exclusive $\eta_c$ production in electron-ion collisions is, in principle, feasible, making the future $eA$ colliders a supplementary source of information about the structure of the pseudoscalar mesons and an important alternative to constrain the $\eta_c$ transition form factor.   


\section{Conclusions}
\label{sec:conc}

In the present letter, we have performed an exploratory investigation of the exclusive $\eta_c$ production in  planned or currently discussed electron-ion
colliders. Assuming that the final state is dominantly produced by $\gamma^* \gamma$ interactions, one has demonstrated that the associated cross-section is determined by the $\eta_c$ transition form factor and, therefore, has its behaviour determined by the description of the meson structure. We have estimated the rapidity, transverse momentum and photon virtuality distributions considering the energy and target configuration expected to be present in the EIC, EicC and LHeC and assuming different potential models to estimate the transition form factor. Our results indicate that future experimental analysis of these distributions is, in principle, feasible and that the associated data can be used to constrain the description of the $\eta_c$ wave function and can supplement the data that can be obtained in $e^+ e^-$ colliders. Such independent data would validate the data obtained
previously by the BaBar Collaboration and allow to provide of detailed tests
of the quarkonium wave function. Our results strongly motivate a more detailed study taking into account the realistic experimental cuts expected for  the detectors proposed to be installed in the EIC at BNL, which we plan to perform in a forthcoming study.

Finally, in this letter, we discussed only $\eta_c$ quarkonium production.
In principle, the analysis can be extended for other final states that can be generated by $\gamma^* \gamma$ interactions. This aspect goes beyond the scope of the present paper, and we leave it for future studies.

\section*{acknowledgments}
V. P. G. would like to thank the members of the Institute of Nuclear Physics Polish Academy of Sciences for their warm hospitality during the initial discussions of this project. V. P. G. was partially supported by CNPq, CAPES, FAPERGS and  INCT-FNA (Process No. 464898/2014-5).
This work was partially supported by the Polish National Science Center grant UMO-2018/31/B/ST2/03537 and by the Center for Innovation and Transfer of Natural Sciences and Engineering Knowledge in Rzesz{\'o}w.

\bibliography{reference.bib}
\end{document}